\documentclass[12pt]{article}
\usepackage{graphicx}
\def\journal{\topmargin 0.0in   \oddsidemargin 0in
        \headheight 0pt \headsep 0pt
        \textwidth 6.5in 
\textheight 9in 
        \marginparwidth 1.5in
        \parindent 2em
        \parskip .5ex plus .1ex         \jot = 1.5ex}
\journal

\def\ra{\rightarrow}
\begin{document}
\begin{titlepage}

\noindent April 2, 2001      \hfill    LBNL-47684\\
Revised October 23, 2001 (for PRL length limit)    \hfill hep-ph/0104024

\begin{center}

\vskip .5in

{\large \bf The $Z \ra \overline bb$ decay asymmetry:\\ lose-lose for the 
Standard Model}
\footnote
{This work is supported in part by the Director, Office of Science, Office
of High Energy and Nuclear Physics, Division of High Energy Physics, of the
U.S. Department of Energy under Contract DE-AC03-76SF00098}

\vskip .5in

Michael S. Chanowitz\footnote{Email: chanowitz@lbl.gov}

\vskip .2in

{\em Theoretical Physics Group\\
     Ernest Orlando Lawrence Berkeley National Laboratory\\
     University of California\\
     Berkeley, California 94720}
\end{center}

\vskip .25in

\begin{abstract}

Combining precision measurements and the Higgs boson search limit, the
electroweak data has evolved to a point where new physics is favored
whether the 3.2$\sigma$ $A_{FB}^b$ anomaly is genuine or not. Such new 
physics could greatly alter the inferred value of the Higgs boson 
mass.

\end{abstract}

\end{titlepage}

\renewcommand{\thepage}{\roman{page}}
\setcounter{page}{2}
\mbox{ }

\vskip 1in

\begin{center}
{\bf Disclaimer}
\end{center}

\vskip .2in

\begin{scriptsize}
\begin{quotation}
This document was prepared as an account of work sponsored by the United
States Government. While this document is believed to contain correct
 information, neither the United States Government nor any agency
thereof, nor The Regents of the University of California, nor any of their
employees, makes any warranty, express or implied, or assumes any legal
liability or responsibility for the accuracy, completeness, or usefulness
of any information, apparatus, product, or process disclosed, or represents
that its use would not infringe privately owned rights.  Reference herein
to any specific commercial products process, or service by its trade name,
trademark, manufacturer, or otherwise, does not necessarily constitute or
imply its endorsement, recommendation, or favoring by the United States
Government or any agency thereof, or The Regents of the University of
California.  The views and opinions of authors expressed herein do not
necessarily state or reflect those of the United States Government or any
agency thereof, or The Regents of the University of California.
\end{quotation}
\end{scriptsize}

\vskip 2in

\begin{center}
\begin{small}
{\it Lawrence Berkeley National Laboratory is an equal opportunity employer.}
\end{small}
\end{center}

\newpage

\renewcommand{\thepage}{\arabic{page}}
\setcounter{page}{1}

{\it Introduction---}A decade of beautiful experiments have provided
increasingly precise tests of the Standard Model (SM) of elementary
particle physics.  The data tests the SM, probes for new physics, and
is sensitive to $m_H$, the mass of the still undiscovered Higgs boson
which gives mass to the elementary particles.  Currently it implies an
upper bound on $m_H$ of order 200 GeV, while direct searches have
established a lower limit of 113.5 GeV.\cite{ewwg}

In the most recent analysis of the electroweak
data the $Z \ra \overline bb$ front-back asymmetry,
$A_{FB}^b$, differs by $3.2 \sigma$ (99.9\% CL) from the SM 
fit.\cite{ewwg}  It could represent new physics, but
a few red flags suggest caution: (1) the direct
determination of $A_b$ from the front-back left-right asymmetry,
$A^b_{FBLR}$, is consistent with the SM ($0.7 \sigma$) while
$A_b$ extracted from $A_b=4A_{FB}^b/3A_l$ (where $A_l$ is the leptonic
asymmetry) disagrees by $3.5 \sigma$ (99.95\% CL), (2) $Z
\ra \overline bb$ measurements have proven notoriously difficult in
the past, and (3) there is no hint of an $R_b$ anomaly to match the
$A_b$ anomaly, requiring a degree of tuning of the left and
right-handed $Z\overline bb$ couplings with an extremely large shift
in the right-handed coupling.

The situation is then quite puzzling. The result could be a
statistical fluctuation, but statistical criteria reviewed below tell
us this is very unlikely.  The remaining two possibilities are new
physics or subtle systematic error.  While great care and effort have
been focused on understanding and reducing the systematic
uncertainties, further work is needed before we can choose clearly
between the two possibilities. If the explanation is systematic error
and $A_{FB}^b$ is omitted, the global SM fit, which is poor with
$A_{FB}^b$ included, becomes excellent, but the predicted value of
$m_H$, the Higgs boson mass, falls to low values in conflict with the
direct search limit, $m_H>113.5$ at 95\% CL.\footnote
{N.B., the 95\% lower limit 
does {\em not} imply a 5\% chance that the Higgs boson is lighter than
113.5 GeV; rather it means that if the Higgs mass were actually 113.5 GeV 
there would be a 5\% chance for it to have escaped detection. The likelihood
for $m_H < 113.5$ GeV from the direct searches is much smaller than
5\%. See for instance the discussion in section 5 of 
\cite{mchvr}.} 
To remove the
inconsistency new physics would be required to modify the predictions
based on the radiative corrections. New physics is then favored
whether $A_{FB}^b$ is affected by systematic error or not, and $m_H$
cannot be predicted without disentangling the effect of the new
physics.

Though less precise it is striking that the values of $x_W^l={\rm
sin}^2\theta^l_W$, the effective leptonic mixing angle, extracted from
the two other hadronic asymmetry measurements, $A_{FB}^c$ and
$Q_{FB}$, agree with $x_W^l$ from $A_{FB}^b$ and deviate from 
the SM fit. Combined the three measurements differ from the SM fit at
99.5\% CL. At the same time they are the only precision measurements
that raise the predicted value of $m_H$ toward the range required by the
direct search limit. 
We show below that all other $m_H$-sensitive precision measurements
favor values much lower than 113.5 GeV.  The set of measurements that are 
consistent with the global fit are inconsistent with the search limit
while the measurements that are essential for consistency with the
search limit are inconsistent with the global fit. 

{\it The data---}In the latest data the $3.5 \sigma$
difference in the SM determinations of $x_W^l$ from $A_{LR}$ and
$A_{FB}^b$ drives a poor fit of the 7 asymmetries used to determine
$x_W^l$, with $\chi^2/dof = 15.5/6$ and CL = 0.013.  The four leptonic
measurements, $A_{LR},A_{FB}^l,A_e,A_{\tau}$, agree very well with one
another, $\chi^2/dof = 2.7/3$, as do the three hadronic determinations
from $A_{FB}^b,A_{FB}^c,Q_{FB}$, $\chi^2/dof = 0.1/2$, while the
aggregated leptonic and hadronic determinations of $x_W^l$ differ by
$3.6 \sigma$ (99.97\% CL).

The four leptonic asymmetries provide the first, third, fourth, and
fifth most precise of the 7 determinations of $x_W^l$. Because they
are consistent, large systematic errors would have to conspire to
affect each measurement in a similar way, which is very unlikely
because they are measured by three very different methods. The same
cannot be said of the hadronic asymmetries, which share common
systematic issues. All three hadronic measurements require similar QCD
corrections and make common use of fragmentation and decay models. As
in the $R_b$ anomaly, $\overline bb$ and $\overline cc$ events
constitute backgrounds for one another.\footnote
{
It is suggestive that the signs of both the $A_{FB}^b$ and $A_{FB}^c$
anomalies are as would be expected if $c$'s were misidentified as
$\overline b$'s and vice-versa, although the systematic error
currently budgeted to this effect is much smaller\cite{ewwg} than
the anomalies.  
}

In the most recent analysis the $A_{FB}^b$ and $A_{FB}^c$ measurements
are assigned a 16\% correlation\cite{ewwg}.

Taking a wider perspective, it is useful to consider the 15 degrees of
freedom in the global SM fit of all data reported in reference
\cite{ewwg}. Even in that framework a $\geq 3.2 \sigma$ discrepancy is
very unlikely, with probability $1 - 0.9986^{15} = 0.021$.  
As noted above, $A_{FB}^b$
also drives the poor $\chi^2$ of that fit, $\chi^2/dof = 26/15$ and CL
= 0.038. With the contribution of $A_{FB}^b$ removed the same fit
parameters yield $\chi^2/dof = 15.8/14$ corresponding to a robust CL =
0.33. If instead the second most deviant measurement, $A_{LR}$, is
omitted, the improvement is much smaller, with $\chi^2/dof = 23.2/14$
and CL = 0.057.

Another feature of the data also points to $A_{FB}^b$ as the `odd man
out.'  The $Z\overline bb$ vertex factor, $A_b$, is predicted very
precisely in the SM and agrees well ($0.7\sigma$) with the direct
determination at the SLC from $A_{FBLR}^b$. But the determination from
$A_b= 4A_{FB}^b/3A_l$ disagrees with the SM by $3.5 \sigma$ and from
the directly measured $A^b_{FBLR}$ by $1.8\sigma$ .

The evidence for new physics in the $Z\overline bb$ vertex is
compelling on a purely statistical level, and the third generation
quarks are a plausible venue for new physics connected to the symmetry
breaking sector. But the disagreement with $A_{FBLR}^b$ and the past
history of $Z\ra \overline bb$ measurements suggest caution.  While
the lessons of the $R_b$ anomaly have been refined and applied to
$A_{FB}^b$, the latter measurement involves 
additional subtleties. Systematic error could in principle provide an
escape path for the SM.  But we will see in the next section that the
path is rather narrow if it is open at all.

{\it Results---}In this section we present $\chi^2$ fits of
$m_H$ and compare them with the search limit. To confront the
predictions of the SM as directly as possible we focus on the directly
measured, $m_H$-sensitive observables.  The observables with the
greatest impact are $x_W^l$ and $m_W$.  The other directly measured,
$m_H$-sensitive $Z$-pole observables are the total width $\Gamma_Z =
2.4952 (23)$ GeV and the ratio of hadronic to leptonic partial widths,
$R_l = \Gamma_h/\Gamma_l = 20.767 (25)$.\cite{ewwg} For $m_Z$, $m_W$
and $m_t$ we use the directly measured values, currently $m_Z =
91.1875 (21)$ GeV, $m_W = 80.448 (34)$ GeV and $m_t=174.3 (5.1)$
GeV.\cite{ewwg}

The strong coupling is taken to be $\alpha_S(m_Z)= 0.118 (3)$.  The
greatest parametric uncertainty is from the electromagnetic coupling
at the $Z$ pole, $\alpha(m_Z)$, in particular from $\Delta \alpha_5$,
the five flavor hadronic contribution to $\Delta \alpha$, which
renormalizes $\alpha$ by $\alpha(m_Z)= \alpha(0)/(1-\Delta
\alpha$). We use five determinations:
two experiment-driven\cite{jd,bp} based on the most
recent data and three theory-driven.\cite{dh,ks,mor}\footnote
{ 
The older theory-driven
determinations\cite{dh,ks} are included because they are consistent
with the new data.  
}  
Gaussian errors are assumed for all experimental quantities. We use
the two loop radiative correction package from ZFITTER\cite{zfitter},
version 6.30, to compute the SM values of the four observables as a
function of $m_H$ and the four experimentally determined parameters,
$m_Z$, $m_t$, $\alpha(m_Z)$, and $\alpha_S(m_Z)$.  Taking as inputs
the all-data fit values\cite{ewwg} for $m_Z$, $m_t$, $\Delta
\alpha_5$, $\alpha_S$, and $m_H$, we reproduce the results (from
ZFITTER v6.35) in \cite{ewwg} as follows: $x_W^l$: 0.23142/0.23142,
$m_W$: 80.394/80.393, $\Gamma_Z$: 2.4960/2.4962, and $R_l$:
20.737/20.740. The effect of such differences on $\chi^2$ is
negligible.

For two reasons we first consider just $m_W$ and
the $Z$-pole measurements, $\Gamma_Z$ and $R_l$: (1) they
are not affected by the issues affecting the asymmetries
and (2) the determination of $m_H$ from them is
less sensitive to the uncertainty from $\alpha(m_Z)$.
The results are striking. Figure 1 shows $\Delta \chi^2 = \chi^2 
- \chi^2_{\rm min}$ as
a function of $m_H$ obtained from $m_W$ alone and in combination with
$\Gamma_Z$ and $R_l$. At each value of $m_H$ the experimental
parameters $m_t$, $\alpha(m_Z)$, and $\alpha_S(m_Z)$ are chosen
to minimize the sum of the $\chi^2$ contributions from $m_W$, $m_t$,
$\alpha(m_Z)$, and $\alpha_S$ (and also from $\Gamma_Z$ and $R_l$ in
the case of the second fit).\footnote
{
We have checked that varying $m_Z$ has negligible effect
on $\chi^2$.
}
The results are summarized in table 1 for
the five choices of $\alpha(m_Z)$. For the fit based just
on $m_W$ the central value of $m_H$ falls between 21 and 28 GeV, with
$m_H < 113.5$ GeV favored at 94 to 92\% CL.  For the fit with
$\Gamma_Z$ and $R_l$ included, the results are even less sensitive to
$\alpha(m_Z)$ and are shifted to lower $m_H$, 15 - 17 GeV, with
CL($m_H < 113.5$ GeV) increased to between 98 and 97\%.

We next consider the effect of the asymmetry measurements in the
framework of the hypothesis that the $A_{FB}^b$ anomaly results from
undetected systematic error.  As discussed above, $A_{FB}^b$,
$A_{FB}^c$ and $Q_{FB}$ share common sytematics so that the most
reliable determination of $x_W^l$ would in this case be provided by
the four leptonic asymmetry measurements, which are very unlikely to
have common systematic uncertainties. Results based on the leptonic
asymmetries, which yield $x_W^l=0.23113\ (20)$, combined with $m_W$,
$\Gamma_Z$, and $R_l$ are shown in figure 2 and summarized in table 2
where they are labeled `$+\ L4$.'  The central values are in the range
27 - 44 GeV, with CL($m_H < 113.5$ GeV) from 98 to 94\%.  As in table
1 there is a significant conflict with the search limit, though with
more dependence on $\alpha(m_Z)$.

For completeness we also consider the effect of the two
lower precision hadronic asymmetry measurements, combining $Q_{FB}$
($x_W^l=0.23118\ (20)$) and $ A_{FB}^c$ ($x_W^l=0.23127\ (19)$)
sequentially with the previously considered observables. The results, in
figure 2 and table 2, conflict with the search limit, though less
decisively. Finally we exhibit the results with all seven asymmetry
measurements included, denoted `$+\ A_{FB}^b$' in table 2, with
$x_W^l=0.23156\ (17)$. As for the usual global SM fits, $m_H$ is
centered around 100 GeV and the fits are consistent with the search
limit for all $\alpha(m_Z)$.

To summarize, each fit with $A_{FB}^b$ omitted is in conflict
with the search limit, and the fits based on the data that is most
reliable if the $A_{FB}^b$ anomaly is a systematic effect --- table 1
and the `$+\ L4$' fit of table 2 --- have the most significant
conflicts.

{\it Discussion---}The greatest source of uncertainty is the
sensitivity to $\alpha(m_Z)$ of the fits that include asymmetry data,
which makes the lack of asymmetry data in table 1 especially
interesting.  The theoretical uncertainty from uncalculated diagrams
is smaller than from $\alpha(m_Z)$, as can be seen in figure 13 of
\cite{ewwg} where the ``blue band'' estimating the theoretical
uncertainty is less than the difference resulting from $\alpha(m_Z)$
for \cite{bp} compared to \cite{mor}.  The figure also shows that the
prediction of ZFITTER as employed in \cite{ewwg} lies near the
large-$m_H$ edge of the blue band. Since, as noted above, our ZFITTER
calculations closely reproduce those of \cite{ewwg}, our estimates of
the conflict with the search limits are conservative relative to the
other libraries/settings used to generate the blue band.

We have found that new physics is favored by the data whether the
$A_{FB}^b$ anomaly reflects new physics or systematic error. An
important consequence is that the evidence from the SM fit favoring a
light Higgs boson becomes less credible. It can be maintained only if
the $A_{FB}^b$ anomaly turns out to be a statistical
fluctuation.\footnote 
{ 
If the discrepancies in the SM fit are
statistical fluctuations, the appropriate fits are those in the bottom
line of Table 2, and the measurements of all the $m_H$-sensitive 
observables, not just
$A_{FB}^b$, must have fluctuated significantly from their true values.
}  
The most generous estimate of the likelihood of this possibility is
the poor 3.8\% $\chi^2$ CL of the global SM fit,\cite{ewwg}, 
which is due almost entirely to the deviation of $A_{FB}^b$ from the 
fit as noted above. Otherwise, whether
the anomaly is a genuine signal of new physics or a systematic
artifact, $A_{FB}^b$ cannot be used to determine $x_W^l$, and the
resulting conflict with the search limit favors new physics
contributions to remove the contradiction. 

We can get a rough idea of the new physics contributions that would be
needed by considering just $x_W^l$ and $m_W$, using the deviation from
the SM for any given value of $m_H$, $\delta x_W^l$ and $\delta m_W$,
to compute the corresponding oblique parameters\cite{pt} $S$ and
$T$. Taking $x_W^l$ from the 4 leptonic asymmetries and using the
direct measurement of $m_W$, we find, e.g., for $m_H= 114,\ 300,\
1000,\ 2000$ GeV that the corresponding values are $(S,T)=$ (-0.02,
0.16), (-0.08, 0.27), (-0.11, 0.48), (-0.09, 0.65), where $m_H= 2000$
GeV is a `stand-in' for dynamical symmetry breaking. The existing data 
cannot choose among these possibilities.

The unexpected emergence of evidence for new physics at the end of the
LEP/SLC decade is a cautionary signal to keep an open mind as to the
ultimate explanation.  If the $A_{FB}^b$ anomaly is genuine, it
signals new physics not anticipated by popular theoretical models. If
the anomaly is genuine and unique to the third generation, it will
also affect $\overline bs$, $\overline bd$, and $\overline sd$ neutral
currents via non-SM $Z$ penguin amplitudes, though the precise effects
are not readily predicted.  If the anomaly is not genuine, the
conflict with the search limit is for now our only evidence of the new
physics and we are left with even fewer clues as to its nature.  

The evidence for new physics presented here may be weakened or
strengthened by future measurements, not only of $A_{FB}^b$ and the
other asymmetries but also of $m_W$ and $m_t$.  New facilities will be
needed to answer the questions posed by the current data, including a
second generation $Z$ factory.  Better measurements of $R_{e^+e^-}$
would be needed to determine $\alpha(m_Z)$ with enough precision to
realize the potential precision of a new $Z$ factory for
$x_W^l$.\cite{jd} This will be important even after the Higgs sector
is discovered, since precise comparisons of the electroweak data with
predictions based on the observed Higgs sector will provide invaluable
guidance on whether additional new physics exists at yet higher
scales.  The evidence of the present data for unspecified new physics
underscores the importance of framing the search for the Higgs sector
in the most general form.

{\it Note added---}Data presented after this work was
completed\cite{summer01} differs slightly from the data considered 
above: $A_{FB}^b$ disagrees with the SM fit by 2.9$\sigma$ (CL
= 99.6\%), and the leptonic and hadronic SM determinations of $x_W^l$
disagree by 3.3$\sigma$ (CL = 99.9\%). The likelihood of the statistical
fluctuation hypothesis increases to a still small probability, e.g.,
from 3.8\% to 6.7\% as gauged by the global fit. The analysis of the
systematic error hypothesis is unaffected since the fits which omit
$A_{FB}^b$ change very little, and the contradiction with the search limit
persists at the levels quoted above.

\vskip 0.2in

\noindent {\bf Acknowledgements} I am grateful to A. H\"ocker,
P. Rowson, and R. Cahn for very helpful discussions. I also wish to
thank T. Abe, D. Bardin, M.  Gr\"unewald, and G.  Passarino for useful
correspondence and H. Chanowitz for computing facilities.

\newpage
\vskip 0.5in

\noindent Table 1. Fit results without asymmetry measurements for five
determinations of $\alpha(M_Z)$.  For each fit the central value of
$m_H$ is shown in GeV and the confidence level (CL) for $m_H>113.5$ GeV.
Results for $m_W$ alone are shown in the first two rows, and for the
combination of $m_W$, $\Gamma_Z$, and $R_l$ in the last two.

\begin{center}
\vskip 12pt
\begin{tabular}{c|c|ccccc}
 && J\cite{jd}& BP\cite{bp} & MOR\cite{mor}& DH\cite{dh} & KS\cite{ks}\\ 
 & $\Delta \alpha_5$   & 0.027896(395) & 0.02761(36)& 0.027426(190)
     & 0.02763(16) & 0.02775(17)  \\
\hline
\hline
$m_W$ & $m_H$ & 21 & 26 & 28 & 25 & 23 \\
& CL & 0.057 & 0.071 & 0.080 & 0.068 & 0.062 \\
\hline 
$+\ \Gamma_Z,R_l$ & $m_H$ & 15&17&15&17& 15 \\
 &CL&0.023&0.028&0.032&0.027&0.024\\
\hline
\hline
\end{tabular}
\end{center}

\vskip 1.2in

\noindent Table 2. Fit results with asymmetry measurements
included. The format is as in table 1. The first fit reflects the
combination of $m_W$, $\Gamma_Z$, and $R_l$ together with the four
leptonic asymmetry measurements, `$L4$'. Successive fits
incrementally include $Q_{FB}$, $A_{FB}^c$, and $A_{FB}^b$.

\begin{center}
\vskip 12pt
\begin{tabular}{c|c|ccccc}
 && J\cite{jd}& BP\cite{bp} & MOR\cite{mor}& DH\cite{dh} & KS\cite{ks}\\
 & $\Delta \alpha_5$   & 0.027896(395) & 0.02761(36)& 0.027426(190)
     & 0.02763(16) & 0.02775(17)  \\
\hline
\hline
$+\ L4$ & $m_H$ &27&37&44&39&34\\
& CL & 0.019&0.041&0.060&0.033&0.023\\
\hline
$+\ Q_{FB}$ & $m_H$ & 33&43&57&46&39 \\
 &CL&0.028&0.058&0.087&0.049&0.034 \\
\hline
$+\ A_{FB}^c$ & $m_H$ & 43&53&61&61&55 \\
 & CL & 0.053&0.10&0.14&0.091&0.069 \\
\hline
$+\ A_{FB}^b$ & $m_H$ &86&110&114&102&89 \\
 & CL & 0.26&0.36&0.50&0.40&0.38 \\
\hline
\hline
\end{tabular}
\end{center}




\newpage

\begin{figure}
\begin{center}
\includegraphics[height=6in,width=4in,angle=90]{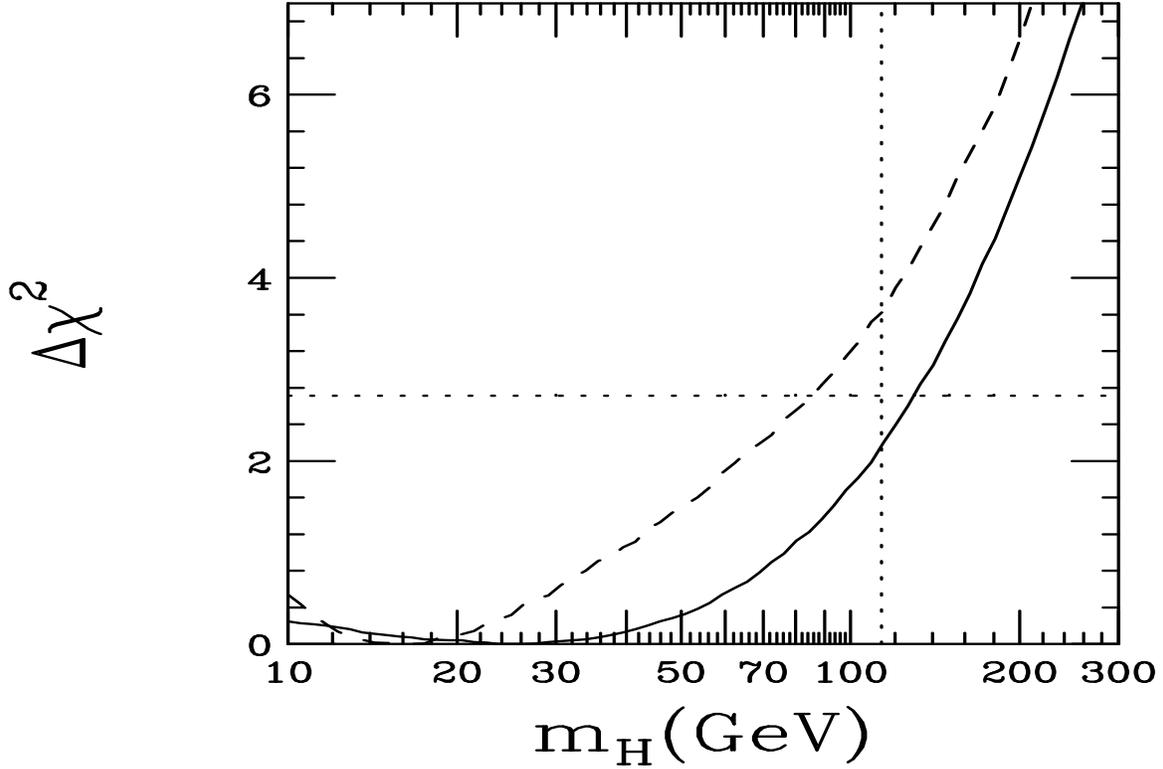}
\end{center}

\caption{$\chi^2$ distributions as a function of $m_H$. The solid line 
is obtained from $m_W$ alone and the dashed line from the combined fit 
of $m_W$, $\Gamma_Z$, and $R_l$. The vertical dotted line indicates the 
direct search lower limit and the horizontal dotted line indicates the 
value of $\Delta \chi^2$ corresponding to a 95\% CL upper limit.
$\alpha(m_Z)$ is from \cite{bp}.}

\label{fig1}
\end{figure}

\newpage

\begin{figure}
\begin{center}
\includegraphics[height=6in,width=4in,angle=90]{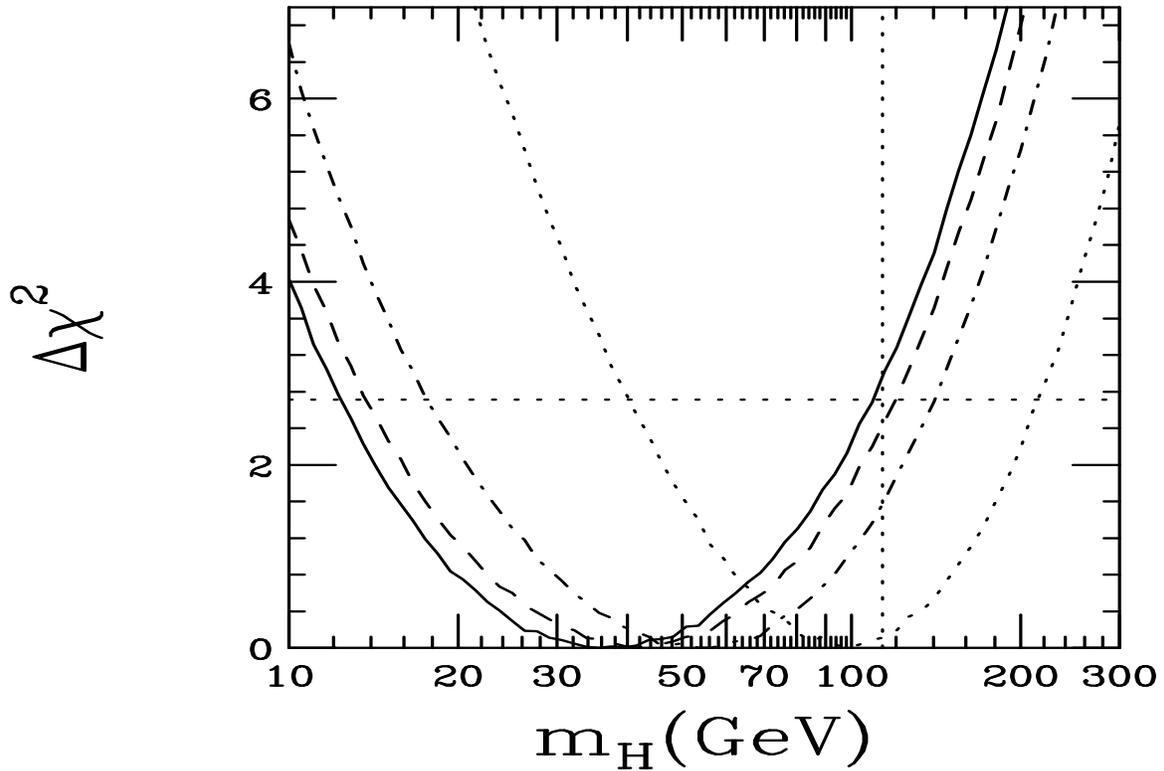}
\end{center}

\caption{$\chi^2$ distributions as in figure 1. The lines correspond
to fits of $m_W$, $\Gamma_Z$, and $R_l$, combined incrementally, as in
table 2, with the four leptonic asymmetry measurements (solid), plus
$Q_{FB}$ (dashes), plus $A_{FB}^c$ (dot-dashes), plus $A_{FB}^b$
(dots). }

\label{fig2}
\end{figure}

\end{document}